\pdfoutput=1
\documentclass{article}
\usepackage{arxiv}

\usepackage[utf8]{inputenc} 
\usepackage[T1]{fontenc}    
\usepackage{hyperref}       
\usepackage{url}            
\usepackage{booktabs}       
\usepackage{amsfonts}       
\usepackage{nicefrac}       
\usepackage{microtype}      
\usepackage{lipsum}		
\usepackage{graphicx}
\usepackage{natbib}
\usepackage{doi}
\usepackage{graphicx}
\usepackage{amsmath}
\usepackage{amssymb}
\usepackage{bm}
\usepackage{amsthm}
\usepackage{booktabs}
\usepackage{algorithm}
\usepackage{algorithmic}
\usepackage[switch]{lineno}
\usepackage{siunitx}  
\usepackage{listofitems} 
\usepackage{tikz}
\usepackage{subfig}
\usepackage{makecell}
\usepackage{multirow}
\usepackage{bbding}
\graphicspath{{figures/}}
\hypersetup{
colorlinks=true,
linkcolor=black,
citecolor=mydarkblue
}

\usetikzlibrary{arrows.meta} 
\usepackage[outline]{contour} 
\contourlength{1.4pt}
\usepackage{pgfmath}
\usepackage{ifthen}
\usepackage{pdfpages}

\usepackage{bm}
\usepackage{xcolor}
\colorlet{myred}{red!80!black}
\colorlet{myblue}{blue!80!black}
\colorlet{mygreen}{green!60!black}
\colorlet{myorange}{orange!70!red!60!black}
\colorlet{mydarkred}{red!30!black}
\colorlet{mydarkblue}{blue!40!black}
\colorlet{mydarkgreen}{green!30!black}
\usetikzlibrary{fit}
\usepackage{orcidlink}
\tikzset{
  >=latex, 
  node/.style={thick,circle,draw=myblue,minimum size=15,inner sep=0,outer sep=0.1},
  node in/.style={node,green!20!black,draw=mygreen!30!black,fill=mygreen!25},
  node hidden/.style={node,black,draw=gray!30!black,fill=gray!40!white,minimum size=10},
  node out/.style={node,blue!20!black,draw=myblue!30!black,fill=myblue!20},
  node pinn/.style={node,red!20!black,draw=myred!30!black,fill=myred!20,minimum size=20},
  connect/.style={thick,mydarkblue}, 
  connect arrow/.style={-{Latex[length=4,width=3.5]},thick,mydarkblue,shorten <=0.5,shorten >=1},
  node 1/.style={node in}, 
  node 2/.style={node hidden},
  node 3/.style={node out}
}
\title{RINN: One Sample Radio Frequency Imaging based on \\Physics Informed Neural Network}

\author{Fei~Shang$^{\orcidlink{0000-0002-5495-8869}}$ \\
	University of Science and Technology of China\\
	\texttt{feishang@mail.ustc.edu.cn} \\
\And
	Haohua Du$^*$$^{\orcidlink{0000-0002-8492-3990}}$ \\
	Beihang University\\
	\texttt{duhaohua@buaa.edu.cn} \\
    \And
	   Dawei Yan$^{\orcidlink{0000-0002-9848-3017}}$ \\
	   University of Science and Technology of China\\
	   \texttt{yandw@mail.ustc.edu.cn} \\
    \And
	   Panlong Yang$^*$$^{\orcidlink{0000-0003-1057-2793}}$ \\
		  Nanjing University of Information Science and Technology\\
		  \texttt{plyang@ustc.edu.cn} \\
 \And
	Xiang-Yang Li$^{\orcidlink{0000-0002-6070-6625}}$ \\
	University of Science and Technology of China\\
	\texttt{xiangyangli@ustc.edu.cn} 
}

\renewcommand{\shorttitle}{\textit{arXiv} Template}

\hypersetup{
pdftitle={A template for the arxiv style},
pdfsubject={q-bio.NC, q-bio.QM},
pdfauthor={David S.~Hippocampus, Elias D.~Striatum},
pdfkeywords={First keyword, Second keyword, More},
}

\begin{document}
\maketitle

\begin{abstract}
    Due to its ability to work in non-line-of-sight and low-light environments, radio frequency (RF) imaging technology is expected to bring new possibilities for embodied intelligence and multimodal sensing. However, widely used RF devices (such as Wi-Fi) often struggle to provide high-precision electromagnetic measurements and large-scale datasets, hindering the application of RF imaging technology. In this paper, we combine the ideas of PINN to design the RINN network, using physical constraints instead of true value comparison constraints and adapting it with the characteristics of ubiquitous RF signals, allowing the RINN network to achieve RF imaging using only one sample without phase and with amplitude noise. Our numerical evaluation results show that compared with 5 classic algorithms based on phase data for imaging results, RINN's imaging results based on phaseless data are good, with indicators such as RRMSE (0.11) performing similarly well. RINN provides new possibilities for the universal development of radio frequency imaging technology.
\end{abstract}

\keywords{EM, RF imaging, Maxwell equations, phaseless data}

\section{Introduction}
Thanks to the characteristic that radio frequency (RF) signals can penetrate targets, RF-based imaging technology has attracted much attention and has been widely used in various fields such as medical imaging and security inspection~\cite{chenWirelessSensingMaterial2024,yanWiPainterFinegrainedMaterial2023}.
RF imaging is based on a fundamental observation: \textit{as an electromagnetic wave, the transmission properties of RF signals depend on the distribution of permittivities in space}~\cite{cho2021mmwall,shang2022liqray}. 
Since different materials have different permittivities, once we can resolve the permittivities at different positions in space from received signals, we have completed RF imaging~\cite{yanWiPainterFinegrainedMaterial2023}. 

The process of inferring permittivities from received signals is called the electromagnetic inverse scattering problem (EISP)~\cite{chen2018computational,peterson1998computational}. 
The EISP designed based on professional equipment and testing environment has long been a focus of researchers.
Based on \textit{precise electromagnetic measurements}, many sophisticated numerical algorithms have been proposed, such as Born approximation~\cite{chew1990reconstruction,devaney1982inversion} and subspace optimization method~\cite{chen2009application,chen2018computational}. 
Due to the strong nonlinearity of inverse scattering problems, many algorithms perform poorly when facing strong scatterers (permittivity significantly greater than 1)~\cite{jinDeepConvolutionalNeural2017}.
In recent years with the development of machine learning technology, many researchers have attempted two-stage schemes~\cite{luoImagingInteriorsImplicit2024,xiaoDualModuleNMMIEMMachine2020} using classical algorithms to obtain an initial solution, and then training a deep learning network on a \textit{large-scale correct labeled radio frequency dataset}~\footnote{For instance, commonly used simulation datasets generated based on MNIST contain thousands of samples.} to enhance the results to achieve high-quality imaging results.

Although these well-known solutions have achieved excellent results, \textit{the dependence on professional equipment and testing environments has hindered the application of EISP technologies in more universal fields}~\cite{caiUbiquitousAcousticSensing2022}. RF signals have non-line-of-sight and weak light sensing capabilities~\cite{chenWirelessSensingMaterial2024}, which are good complements to vision-based sensing schemes. In the real world, there are hundreds of billions of ubiquitous RF devices (approximately 21.1 billion Wi-Fi devices alone~\cite{wifi-device}), which cannot provide high-precision electromagnetic measurements (for example, the phase data is directly randomly distributed in $[0,2\pi]$~\cite{hanACEAccurateAutomatic2020,terryOFDMWirelessLANs2003}) and lack large-scale datasets~\footnote{Although RF data is abundant, it cannot be as easily annotated with accurate information as simulation data based on MNIST, resulting in a limited number of available samples.}~\cite{chenWirelessSensingMaterial2024}. If we can use machine learning technology to transform the EISP problem so that it can obtain imaging results based on a small number of samples (even one sample ) and low-quality data, this will greatly expand the potential applications of existing multimodal sensing technologies, including post-disaster search, health monitoring, etc.

We re-examine the EISP: we expect to obtain a set of solutions (permittivity coefficients) that satisfy the transmission characteristics of electromagnetic waves (Maxwell's equations).
The concept of embedding physical information into neural networks has been extensively applied in scientific discovery~\cite{matthewsPinnDEPhysicsInformedNeural2024}. We posit that it also holds significant potential for enhancing the universality of imaging technologies through sensing.
If we can leverage physics-informed neural networks (PINNs) ~\cite{raissiPhysicsinformedNeuralNetworks2019} to embed the transmission characteristics of electromagnetic waves into the neural network as prior knowledge, guiding and constraining its convergence, we can significantly reduce the search space in the inversion of radio frequency (RF) data. This enables the estimation of the spatial distribution of complex permittivity coefficients using one sample. Furthermore, if the network can be adapted to handle low-quality data, it may open the possibility of achieving imaging using common RF communication devices such as Wi-Fi and mobile phones.
By inferring the distribution of complex permittivity coefficients in space from only one sample data, we no longer need large-scale professional data for training; if we can further adapt the network to low-quality data, we have the opportunity to achieve imaging on common RF communication devices such as Wi-Fi and mobile phones. 

In this paper, we propose \textbf{RINN}, a neural network framework that can handle phaseless data, leveraging physical constraints instead of large-scale datasets to achieve network convergence and perform RF imaging.
However, there are two key challenges that need to be addressed first.
(1) The inherently ill-posedness and nonlinearity of the inverse scattering problem of electromagnetic signals make it difficult to solve~\cite{jinTheoryComputationElectromagnetic2015}. The discretization in the numerical computation process brings sampling errors, reduces image resolution, and exacerbates the difficulty of distinguishing targets at close distances~\cite{xiaoDualModuleNMMIEMMachine2020}. (2) Although the lack of phase information in data does not destabilize the solution, it can worsen instability~\cite{ammariPhasedPhaselessDomain2016}. Even worse, we hope that using one sample for training will lead to network convergence, which poses a serious risk of overfitting.

Many works~\cite{weiDeepLearningSchemesFullWave2019,xiaoDualModuleNMMIEMMachine2020} estimate only the difference in permittivities. We use Implicit Neural Representations (INR) to simultaneously estimate induced currents and permittivities, which avoids the nonlinearity and pathological issues that may arise from matrix inversion. Furthermore, previous research has shown that the flexibility of INR in handling image resolution also helps alleviate sampling errors caused by discretization~\cite{chen2021learning,cheng2024colorizing}. Additionally, we view the INR network as a mapping between two random variables rather than numerical fitting, introducing randomness to mitigate ``population assumption" problems~\cite{song2020score}, reduce overfitting on one sample data, and enhance stability.

Our main contributions can be summarized as follows: 
(1) We have designed \textbf{RINN} by introducing physical constraints, which enables imaging based on one sample and phaseless data, making it possible to deploy EISP technology on billions of ubiquitous RF devices.
(2) From the perspective of random variable coding, we have designed a neural network that improves the ill-conditioning and instability issues caused by matrix inversion in traditional solutions and training with one sample data.
(3) Our numerical results indicate that the \textbf{RINN} network based on phase less data imaging performs well compared to five well-known algorithms based on phase data imaging, with levels of metrics such as RRMSE (0.11) being close.

\section{Related work}
\label{sec:related}
\textit{Electromagnetic inverse scattering problem.}
As a classic problem, the inverse scattering problem has attracted extensive attention.
Traditional solution approaches mainly have two perspectives. 
(1) Based on certain assumptions, approximate and simplify equations, and then solve them. Classical algorithms include Born approximation~\cite{paganoChallengesOperationalWeather2024}, Rytov approximation~\cite{devaneyInversescatteringTheoryRytov1981}, etc. However, the applicability of such algorithms is limited by the approximations made; for example, Born approximation is only applicable when the complex permittivity of scatterers is close to that of air. (2) Using iterative schemes to optimally estimate the problem; classical algorithms include subspace methods~\cite{chen2018computational}. 
However, these algorithms are highly dependent on initial solutions~\cite{ammariPhasedPhaselessDomain2016}. 
In addition, classical algorithms often rely on phase information of electric fields and are not suitable for phaseless data~\cite{chen2018computational}
In recent years, with the development of machine learning technology, many excellent researchers~\cite{luoImagingInteriorsImplicit2024,xiaoDualModuleNMMIEMMachine2020,chenREVIEWDEEPLEARNING2020,jinDeepConvolutionalNeural2017} have attempted to use neural networks to solve electromagnetic inverse scattering problems. They usually consist of two stages, in the first stage using classical methods to obtain an initial solution, and then training an enhancement network with a dataset to improve imaging effects. However, collecting a complete dataset and determining the actual complex permittivities of various targets is difficult in many scenarios~\cite{caiUbiquitousAcousticSensing2022,yangSLNetSpectrogramLearning2023}.


\textit{Physics-informed neural networks}.
The physics-informed neural network (PINN) model classically consists of a deep neural network, characterized by the inclusion of a physics-informed term in the loss function, representing the physical laws to be followed~\cite{matthewsPinnDEPhysicsInformedNeural2024,yangBPINNsBayesianPhysicsinformed2021,zhengImplementationMaxwellsEquations2024}.
For instance, in fluid dynamics, the Navier-Stokes equations might be used as the physical information.
During training, the model not only minimizes data error but also reduces the error associated with the physics-informed term, ensuring that the predictions adhere to physical laws.
In traditional machine learning approaches, the learning process is primarily data-driven, with models heavily relying on large amounts of high-quality data~\cite{zhang2023dive}.
But in practical applications, data scarcity or noisy data often pose significant challenges~\cite{chenWirelessSensingMaterial2024,shang2022liqray}.
Under such circumstances, data-driven models alone struggle to produce accurate and reliable predictions~\cite{wangWiMeasureMillimeterlevelObject2023,yanWiPainterFinegrainedMaterial2023}.
In contrast, PINNs incorporate physical knowledge as a prior, aiming to overcome the limitations of insufficient data.
By leveraging physical laws, PINNs can provide physically intuitive predictions even when data are limited~\cite{matthewsPinnDEPhysicsInformedNeural2024}.

\section{Field sensing model}
\label{sec:model}

\subsection{The ideal model derived from Maxwell's equations}

\begin{figure}
\centering
    \includegraphics[width=0.5\linewidth]{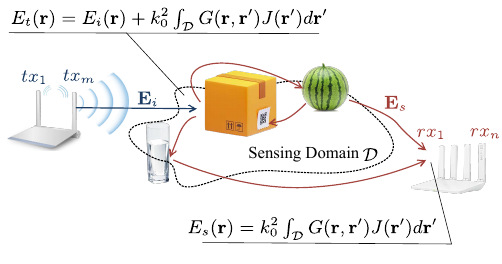}
    \caption{The incident wave $\bm{E}_i$ induces a sensing current $\bm{J}$ in the dielectric medium, thereby exciting scattered waves $\bm{E}_s$. The variation in permittivity among different media provides an opportunity for imaging utilizing these scattered waves.}
    \label{fig:field}
    \vspace{-1em}
\end{figure}

We consider a scenario as depicted in Fig.~\ref{fig:field}. There are \(m\) independent transmitters, with the position of the \(i\)-th transmitter denoted as \(\bm{r}_{tx_i}\); and there are \(n\) independent receivers, with the position of the \(i\)-th receiver denoted as \(\bm{r}_{rx_i}\). The domain of interest, \(\mathcal{D}\), contains several targets of different positions, shapes, and materials.
Due to the varying complex permittivities of different materials, the combination of sensing targets leads to a distinct distribution of the complex permittivity within \(\mathcal{D}\), denoted as \(\varepsilon(\bm{r})\).
We consider a transverse magnetic (TM) wave with a vacuum wavenumber \(k_0\) and only a \(z\)-axis component, assuming that the targets within domain \(\mathcal{D}\) have uniform material properties and a height much greater than their base width, allowing us to approximate them as being homogeneous in the \(z\) direction.
In this case, the wave in space satisfies\cite{peterson1998computational} 
\begin{equation}
    \frac{\partial^2 E}{\partial x^2} + \frac{\partial^2 E}{\partial y^2} + k^2 E = 0,
    \label{eq:maxwell}
\end{equation}
where $E$ is the electric field, $(x,y)$ is the position, and $k$ is the wavenumber that is related to the complex permittivity.
The solution is: 
\begin{equation}
    \left\{
        \begin{aligned}
            E_t(\bm{r}) &= E_i(\bm{r}) + k_0^2 \int_{\mathcal{D}} G(\bm{r}, \bm{r'}) J(\bm{r'}) \quad \text{for } \bm{r} \in \mathcal{D}\\
            E_s(\bm{r}) &= k_0^2 \int_{\mathcal{D}} G(\bm{r}, \bm{r'}) J(\bm{r'})  d\bm{r'} \quad \text{others},
        \end{aligned}\right.
        \label{eq:model}
\end{equation}
where \(E_t(\bm{r})\) and \(E_s(\bm{r})\) are the total and scattered electric fields, respectively; \(E_i(\bm{r})\) is the incident electric field; \(G(\bm{r}, \bm{r'})\) is the Green's function~\cite{bowman1987electromagnetic}; and \(J(\bm{r'})\) is the current density, which is related to the permittivity \(\varepsilon(\bm{r})\) by~\cite{peterson1998computational}
\begin{equation}
    J(\bm{r'}) = [\varepsilon(\bm{r})-1]E_t(\bm{r}).
    \label{eq:current}
\end{equation}

The two integrals in Equ.~\ref{eq:model} indicate that when the Green's function $G(.)$ is relatively large, the complex permittivities at different positions cannot be considered independent in their influence on RF signals.
This also explains why waves no longer travel in straight lines but exhibit diffraction phenomena when the target size is no longer much larger than the wavelength.
Once the distribution of complex permittivities in space is inversely deduced from RF signals, achieving integrated sensing of target imaging and material recognition becomes less challenging: different complex permittivities correspond to different materials, and their boundaries correspond to shape information.

\subsection{Discretization}
In order to facilitate the computation of computer systems, we discretize the above model.
Specifically, we discretize the sensing domain $\mathcal{D}$ into $N$ subregions, with the center coordinates of the ith subregion being $\bm{r}_{d_i}$. 
At this point, Equ.~\ref{eq:model} can be rewritten as:
\begin{equation}
    \left\{
        \begin{aligned}
            \bm{E}_t &= \bm{E}_i + \bm{G}_{\mathcal{D}}\bm{J}\quad \text{for } \bm{r} \in \mathcal{D}\\
            \bm{E}_s &= \bm{G}_{s} \bm{J}\quad \text{others},
        \end{aligned}\right.
        \label{eq:model-d}
\end{equation}
where $\bm{E}_t$ is the total electric fields in the sensing domain $\mathcal{D}$, $\bm{E}_s$ is the scattered fields in the received antennas,  $\bm{E}_i$ is the incident electric field in the sensing domain $\mathcal{D}$. $\bm{G}_{\mathcal{D}}$ and $\bm{G}_{s}$ are the Green's functions corresponding to the sensing domain $\mathcal{D}$ and the rest of the space, respectively; and $\bm{J}$ is the current density vector, which is related to the permittivity $\varepsilon(\bm{r})$ by
\begin{equation}
    \bm{J}[i] = [\varepsilon(\bm{r}_{d_i})-1]\bm{E}_t[i].
    \label{eq:current-d}
\end{equation}

\subsection{Practical issues}
\label{sec:practical}
Previously, numerous artistic works~\cite{xiaoDualModuleNMMIEMMachine2020,chenREVIEWDEEPLEARNING2020,jinDeepConvolutionalNeural2017} have explored EISP; however, when applying these to sensing with ubiquitous RF devices such as Wi-Fi, several practical issues still need to be carefully addressed.

(1) The electric field values required for solving many equations are difficult to measure.
Solving Equ.~\ref{eq:model} requires first obtaining two electric field quantities: the scattered field \( E_s \) at the receiver and the incident field \( E_i \) in the sensing domain $\mathcal{D}$.
However, both quantities are difficult to measure directly. The difficulty in measuring \( E_s \) stems from the fact that the RF signal received is the result of the superposition of the incident and scattered waves, followed by processes such as antenna gain and automatic gain control, rather than \( E_s \) itself~\cite{yangArtificialIntelligenceenabledDetection2022}. Measuring \( E_i \) is even more challenging, as deploying a large number (this depends on the size of the discrete grids, often thousands or more) of receivers in the sensing domain $\mathcal{D}$ to collect signals is often impractical.

(2) WiFi data struggle to provide accurate phase estimates.
It is noteworthy that all variables appearing in the equation are complex numbers.
Due to the lack of precise clock synchronization methods, CSI data contain significant phase noise, such as CFO (Carrier Frequency Offset) and SFO (Sampling Frequency Offset)~\cite{terryOFDMWirelessLANs2003,hanACEAccurateAutomatic2020}.
The poor quality of phase data exacerbates the difficulty in estimating the complex permittivity.

(3) Classical algorithms do not fully meet the requirements of sensing scenarios.
Specifically, model-driven approaches struggle to converge when the complex permittivity of the medium is significantly higher than that of air due to the nonlinearity factors in Equ.~\ref{eq:maxwell} \cite{ammariPhasedPhaselessDomain2016}. For instance, the complex permittivity of common liquids like water (approximately 78~\cite{kaatzeComplexPermittivityWater1989,shangContactlessFineGrainedLiquid2024}) is often dozens of times greater than that of air (approximately 1). Although data-driven methods can significantly enhance inversion performance through the nonlinearity of neural networks, collecting sufficient radio frequency datasets is costly (test equipment can cost hundreds of thousands of dollars~\cite{dhekne2018liquid}).
\begin{figure}[t]
	\subfloat{\includegraphics[width=0.5\linewidth]{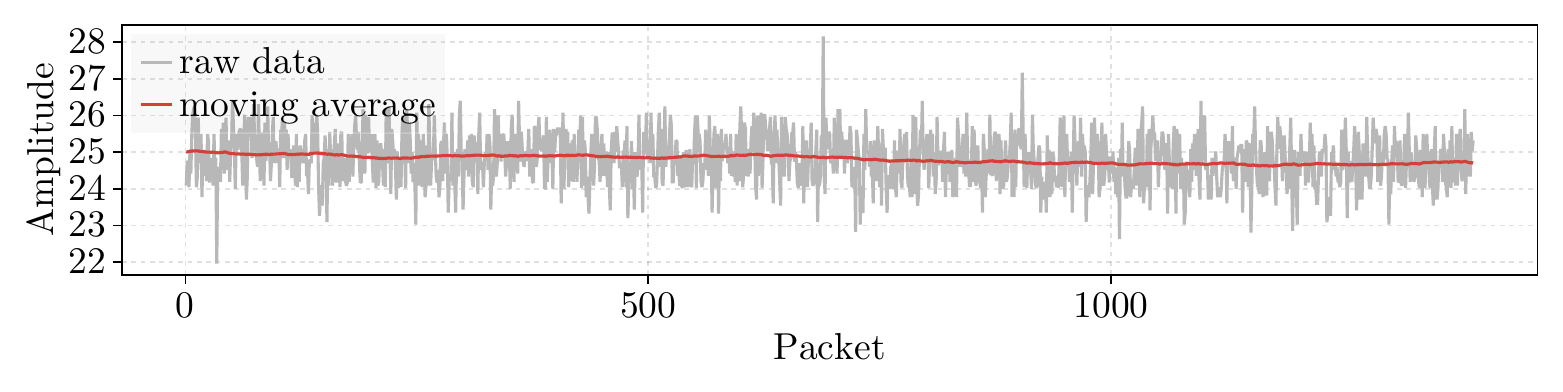}}
    \hfill
	\subfloat{\includegraphics[width=0.5\linewidth]{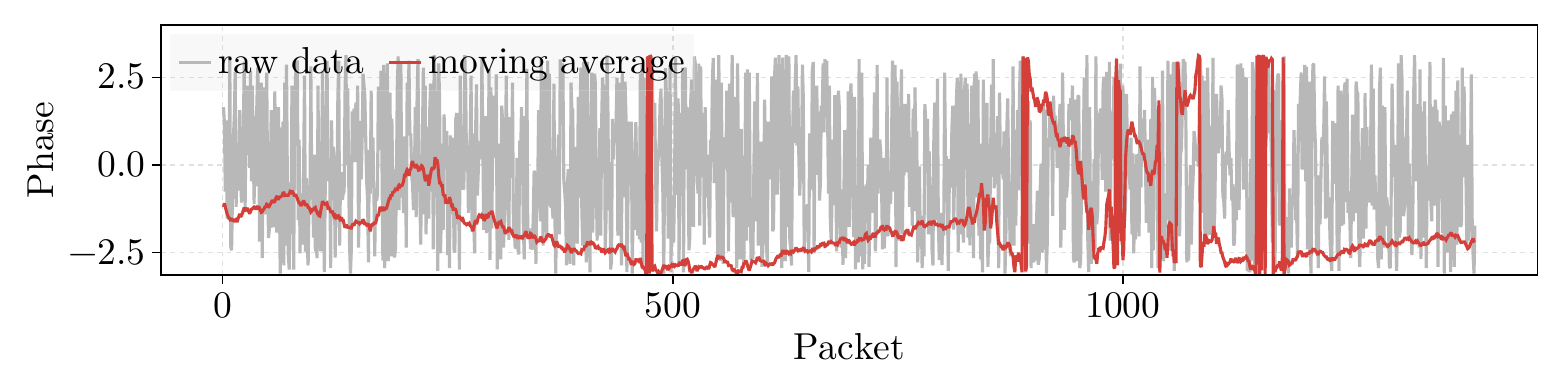}}
	\caption{The amplitude of WiFi signals is more stable than the phase after smoothing. }
	\label{fig:signal}
\end{figure}

\section{System design}
\label{sec:system}


In contrast to the classical EISP problem, to accommodate radio frequency technology, we first need to preprocess the data (Sec.~\ref{sec:preprocessing}) so that it can be used for solving.
Furthermore, leveraging the concept of PINN, we embed physical information into neural networks and design the \textbf{RINN} network (Sec.~\ref{sec:network}). This approach avoids the collection of extensive training data while harnessing the powerful nonlinear capabilities of neural networks, enabling imaging of the sensing domain based on phaseless data.

\subsection{Preprocessing}
\label{sec:preprocessing}
To utilize Wi-Fi data for imaging in field sensing models, we need to address two primary issues: \textbf{data noise} and \textbf{gain differences}. For the amplitude data, which is relatively stable and suitable for our subsequent tasks, we focus on reducing noise by applying mean filtering and removing outliers from the raw data. Since our system employs multiple independent transmitters and receivers, these devices might introduce varying signal gains~\cite{xiePrecisePowerDelay2019}. 
Handling these gain differences is crucial for ensuring reliable imaging.

\begin{figure}
    \centering
    \includegraphics[width=\linewidth]{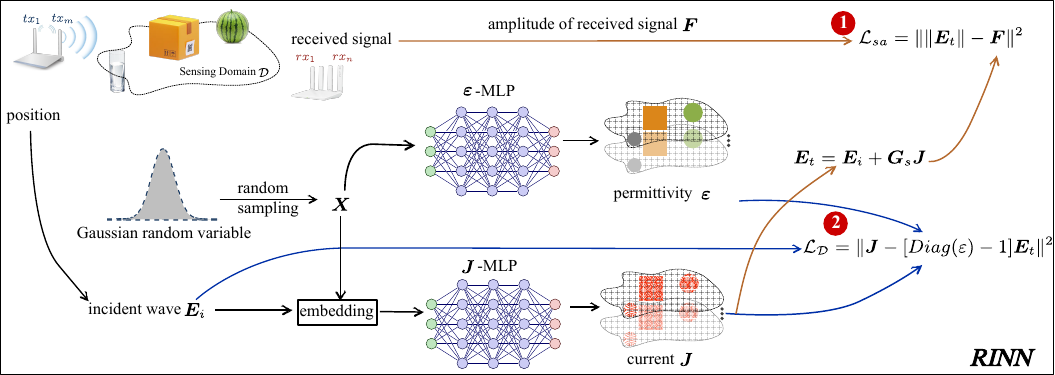}
    \caption{\textbf{RINN}. We use the idea of PINN to design a neural network \textbf{RINN} for imaging based on one sample using phaseless data. The main idea is to obtain a set of solutions that satisfy the constraints of Maxwell's equations (Equ.~\ref{eq:model-d}) with the help of the encoding capability of MLP networks. The uniqueness of the solution to the electromagnetic inverse scattering problem has been proven. We sample from a multidimensional Gaussian distribution to obtain $\bm{X}$, then encode it using $\varepsilon$-MLP to obtain the distribution of complex permittivity; noting that sensing current $\bm{J}$ is a function of incident waves and spatial coordinates, we embed incident waves into $\bm{X}$ and then encode them using $\bm{J}$-MLP network to obtain an estimate $\bm{J}$ for sensing current. Equ.~\ref{eq:model-d} has two constraints: one is that received signals should be as close as possible to actual values, and another is that two representation schemes for $\bm{J}$ are equivalent (outputs from J-MLP and results from Equ.~\ref{eq:current-d}). Based on this we designed loss functions $\mathcal{L}_{sa}$ and $\mathcal{L}_\mathcal{D}$. Then through multiple rounds training until convergence was achieved by our network.}
    \vspace{-1em}
    \label{fig:RINN}
\end{figure}
The received signal $R$ is the result of electromagnetic waves transmitted in space after undergoing processes such as antenna gain and automatic gain control (AGC)~\cite{xiePrecisePowerDelay2019},.
Additionally, it is subject to zero-mean Gaussian noise~\cite{messerEnvironmentalMonitoringWireless2006} which can be represented as:
\begin{equation}
    R = G_{agc} G_{ant} E_t + N,
\end{equation}
where \(G_{agc}\) and \(G_{ant}\) are the AGC gain and antenna gain, respectively; \(E_t\) is the electric field; and \(N\) is the noise.

Since Gaussian noise has a zero mean, we use the expectation of the received signal for subsequent processing.
Fig.~\ref{fig:signal} shows the signal after taking the expectation, revealing that the amplitude exhibits good stability.
However, due to the presence of modulus operation, if the phase error range of the device itself is large, the expected value is not stable.
Therefore, in the subsequent processing, we only use the amplitude of the received signal.
Moreover, the AGC $G_{agc}$ of Wi-Fi devices can be obtained.
Therefore, the issue now lies solely in how to eliminate the impact of antenna gain \( G_{ant} \) differences.
Fortunately, \( G_{ant} \) is only related to the device itself, and we can obtain it through a preliminary calibration, which is:
\begin{equation}
    G_{ant} = \mathbb{E} \left[\frac{R_{w/o}}{G_{agc} E_t}\right],
\end{equation}
where \( R_{w/o} \) is the received signal without the target, and \( \mathbb{E} \) denotes the expectation.
Since there is no scattered wave from the target at this time, the total electric field is equivalent to the incident wave.
According to the propagation characteristics of electromagnetic waves~\cite{chen2018computational}, it is:
\begin{equation}
    E_t = E_i = \frac{e^{j\omega t}}{r},
    \label{eq:ei}
\end{equation}
where \( r \) is the distance between the transmitter and the receiver and \( \omega \) is the angular frequency.

\subsection{RINN}
\label{sec:network}

Unlike many artistic two-stage schemes, we do not have a large amount of labeled data for model training. 
Instead, we use the idea of PINN to try to train a regression network so that the output can satisfy Equ.~\ref{eq:model-d} for the current input signal. 
In this way, we achieve ``one sample sensing".
However, due to having only one sample of input data, it is very easy for the network to overfit.
Inspired by research~\cite{song2021scorebasedgenerativemodelingstochastic}, we view the network as a mapping between two random variables rather than between two numerical values.
The overall framework of \textbf{RINN} is illustrated in Fig.~\ref{fig:RINN}.
Specifically, we designed two multilayer perceptrons (MLP), whose outputs represent the mean values of complex permittivity and induced current, and then used Equ.~\ref{eq:model-d} as a physical constraint to achieve regression.
As a result, we obtain the solution under the current received signal conditions.

\subsubsection{$\varepsilon$-MLP}
We use $\bm{\varepsilon}$-MLP with parameters $\theta_{\varepsilon}$ to map the multidimensional gaussian random variables $\bm{X}$ to the complex permittivity in sensing domain $\mathcal{D}$.
We assume that the output $\bm{\varepsilon}$ of the network is a Gaussian random variable, i.e., $\bm{\varepsilon} \sim \mathcal{N}(\mu_{F_\varepsilon},\Sigma_{\varepsilon})$. 
We use the $\varepsilon$-MLP network to represent a probability mapping $p_{\theta_{\varepsilon}}(\varepsilon|x)$.
The output of the $\varepsilon$-MLP is:
\begin{equation}
    \bm{F_\varepsilon} = F_{\varepsilon}(\bm{X};\theta_{\varepsilon}).
    \label{eq:epsilon-mlp}
\end{equation}


\subsubsection{J-MLP}
After obtaining the complex permittivity \(\bm{\varepsilon}\), the induced current \(\bm{J}\) can be obtained by solving Equ.~\ref{eq:model-d} and Equ.~\ref{eq:current-d} simultaneously, which subsequently allows for the derivation of the scattered wave \(E_s\).
However, this process involves matrix inversion, which is computationally intensive and poses challenges for gradient calculation.
Therefore, we designed a $\bm{J}$-MLP network $F_J$ to obtain the induced current \(\bm{J}\) without performing matrix inversion.
Similar to $\varepsilon$-MLP, $\bm{J}$-MLP is also not a one-to-one mapping of numbers. 
We assume its output $\bm{J}$ is a Gaussian random variable, that is, $J \sim \mathcal{N}(\mu_{F_J},\Sigma_J)$.
$\bm{J}$-MLP acts as an encoder to represent a probability mapping $p_{\theta_J}(J|x,E_i)$.
The output of the $\bm{J}$-MLP is:
\begin{equation}
    \bm{F_J} = F_{J}(\bm{X},\bm{E}_i;\theta_{J}),
    \label{eq:J-mlp}
\end{equation}
where $\bm{E}_i$ is the incident wave in the sensing domain $\mathcal{D}$ which is given by Equ.~\ref{eq:ei}, $\bm{X}$ is the spatial coordinates of the sensing area $\mathcal{D}$, and $\theta_{J}$ are the parameters of the $\bm{J}$-MLP.

\subsubsection{Loss function}
When the network converges, we aim to: (1) make the scattered wave corresponding to the complex permittivity as close as possible to the actual measured data, and (2) ensure that the induced current in the sensing domain $\mathcal{D}$ satisfies the constraints of equation Equ.~\ref{eq:model}.
To achieve these goals, we have designed a \emph{scatter loss} $\mathcal{L}_{sa}$ and a \emph{sensing domain loss} $\mathcal{L}_\mathcal{D}$ to constrain each part, addressing the practical issues mentioned in Sec.~\ref{sec:practical}.

\textit{Scattering loss.} 
An intuitive design is to multiply the output of the J-MLP—induced current \( \bm{F}_{J} \)—with \( G_s \) to obtain the scattered wave, then extract the corresponding scattered signal from the received signal to compute their MSE loss.
However, due to the lack of phase data and the fact that the received signal is a superposition of the incident and scattered waves, we define the scattering loss as:
\begin{equation}
    \begin{aligned}
         \mathcal{L}_{sa} &= \left| \Vert \bm{G}_s \bm{\mu}_{J} + \bm{E_i^R} \Vert - \Vert \hat{\bm{E}}_t \Vert \right|_2\\
            &= \mathbb{E}\left(\left| \Vert \bm{G}_s \bm{F_J} + \bm{E_i^R} \Vert - \Vert \hat{\bm{E}}_t \Vert \right|_2\right)
    \end{aligned}
    \label{eq:scatter}
\end{equation}
where \( \bm{E_i^R} \) is the incident wave in the receivers,  \( \hat{\bm{E}}_t \) is the received signal after preprocessing, $\Vert \cdot \Vert$ is the amplitude operation,  \( | \cdot |_2 \) is the Euclidean norm, and $\mathbb{E}$ means expectation.

\textit{Sensing domain loss.}
To train the $\varepsilon$-MLP, we designed a perception domain loss.
Specifically, the induced current \( \bm{J} \) has two representations, given by Equ.~\ref{eq:model-d} and Equ.~\ref{eq:J-mlp}.
Theoretically, these two should be identical.
Therefore, \( \mathcal{L}_{\mathcal{D}} \) is designed as follows:
\begin{equation}
    \begin{aligned}
        \mathcal{L}_{\mathcal{D}} &= \left| \bm{\mu_J} - [Diag(\bm{\mu_\epsilon})-1](\bm{E}_i + \bm{G}_\mathcal{D}\bm{\mu_J})\right|_2\\
        &= \mathbb{E}\left(\left| \bm{F_J} - [Diag(\bm{F_\epsilon})-1](\bm{E}_i + \bm{G}_\mathcal{D}\bm{F_J})\right|_2\right)
    \end{aligned}
    \label{eq:sensing}
\end{equation}
where \( Diag(\bm{F_\epsilon}) \) is a diagonal matrix with the elements of \( \bm{F_\epsilon} \) on the diagonal.

\textit{Regularization loss function.} Noting that the distribution of materials in space is often continuous (with only abrupt changes occurring at the boundaries of different materials), we embed this physical information into the training process to enhance sensing results. 
Specifically, we use total variation loss~\cite{rudin1992nonlinear} $\mathcal{L}_{tv}$ commonly used in image smoothing to regularize the results.

As  a result, the total loss function is:
\begin{equation}
    \mathcal{L} = \mathcal{L}_{sa} + \lambda_1 \mathcal{L}_{\mathcal{D}} + \lambda_2 \mathcal{L}_{tv}.
    \label{eq:total}
\end{equation}

\subsection{Implementation}
We use Julia scripts for data processing, and implement the \textbf{RINN} using the Lux.jl library~\cite{pal2023efficient,pal2023efficient}.
During training, the sensing sensing $\mathcal{D}$ is discretized into 4096 sub-domains of size 64 $\times$ 64. 
We set the dimension of $\bm{X}$ to be 20, then sample it from a standard normal distribution, with each batch size set to 32. The $\varepsilon$-MLP network has 4 layers, with each hidden layer containing 128 variables. The activation function for the first three layers is ReLU, and for the last layer is sigmoid. We multiply the network output by a scaling factor to estimate the complex permittivity coefficient. The purpose of setting the scaling factor is to ensure that the network output interval contains actual values. The structure of the first three layers in $\bm{J}$-MLP network is similar to $\varepsilon$-MLP, except that there is no activation function in the last layer and two dimensions are present - representing real and imaginary parts of induced current respectively which are combined into a complex number for subsequent calculations.
 We use Adam optimizer with default values, learning rate of $5\times 10^{-4}$, decayed during optimization process following an exponential scheduler.
 We set $\lambda_1$ to 1, $\lambda_2$ to 0 for the first 5000 rounds of training, and then set it to 1 for later training.
 We train the \textbf{RINN} network on NVIDIA GTX4090 GPU.

\section{Evaluation}
\label{sec:eva}
\subsection{Experimental Setup}
\textit{Dataset}. We selected two types of scatterers for testing: one is the ``Austria" pattern commonly used in classical electromagnetic inverse scattering problems, which consists of two small circles and one large ring; the other is the MNIST handwritten dataset commonly used in deep learning electromagnetic inverse scattering problems, from which we randomly selected 1000 images (100 images per handwritten digit). We set the complex permittivity to a random number between 1 and 2. We defined the scattering region \( \mathcal{D} \) as a square with a side length of 32 cm, discretized into \( 64 \times 64 \) subregions. We arranged 16 transmitting antennas and 32 receiving antennas. The transmitting antennas are uniformly distributed on a circle centered at the center of \( \mathcal{D} \) with a radius of 1 m, while the receiving antennas are uniformly distributed on a circle centered at the center of \( \mathcal{D} \) with a radius of 0.5 m. The gain of the transmitting antenna and the receiving antenna is set to 3 dB. We assumed that the signal amplitude excited by the transmitting antennas is 1 and used the method of moments to calculate the scattered signals. The signal calculations were performed using Julia programming language.

\begin{figure}
    \centering
    \subfloat{\includegraphics[width=\linewidth]{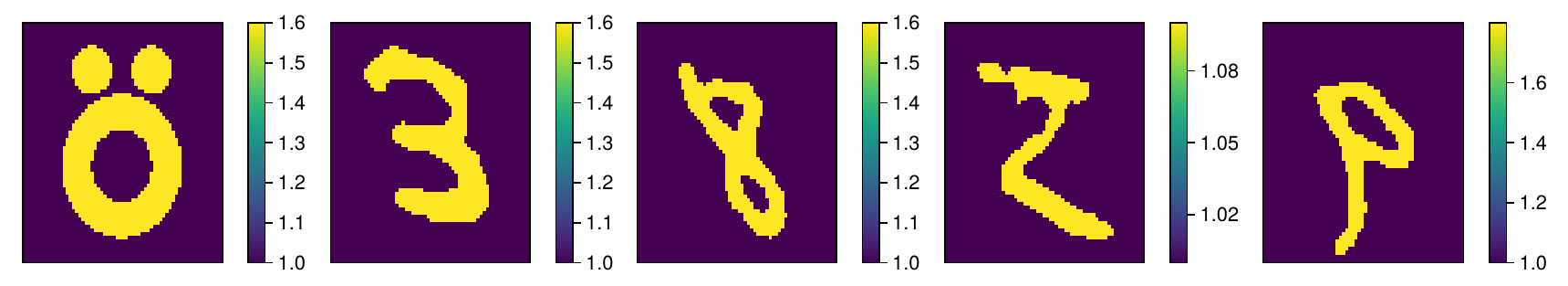}}

    \subfloat{\includegraphics[width=\linewidth]{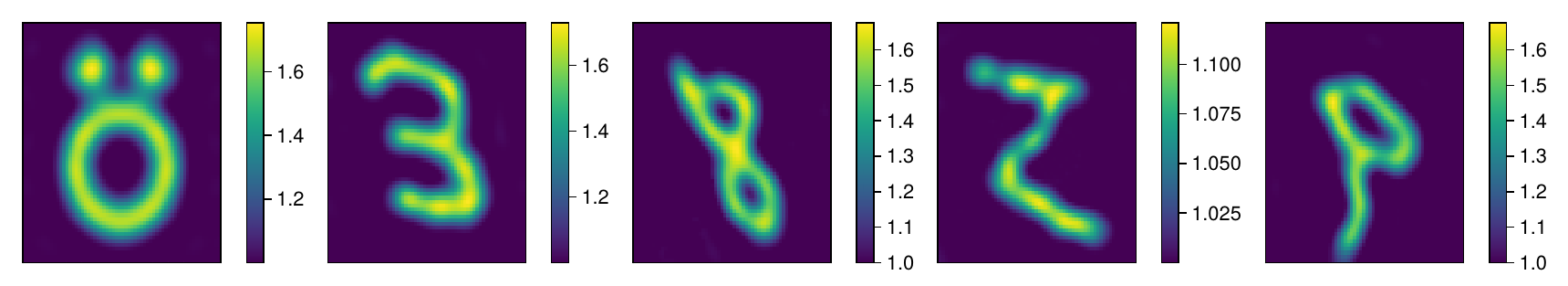}}
    \caption{The imaging results of \textbf{RINN}. We used Austrian patterns and handwritten images from the MNIST dataset as imaging targets, setting them to different permittivities, and then obtained scattering signals using the method of moments simulation. The top row of images shows the ground truth of the targets, while the bottom row shows the imaging results obtained by the \textbf{RINN} network based on phaseless data. The permittivity for the first three columns on the left is set to 1.6, for the fourth column it is set to a smaller value of 1.1, and for the rightmost column it is set to a larger value of 1.8.}
    \label{fig:step}
 \end{figure}
 
\begin{table}
    \centering
    \caption{A comparison of the performance of imaging results from different algorithms. This includes two classical algorithms (BP and GS SoM) and three deep learning approaches (CS-Net, DConvNet, and Physics-Net). Since the \textbf{RINN} is compatible with both phase data and phaseless data, it is evaluated using three metrics: SSIM, PSNR, and RRMSE. In evaluations with phase data, the \textbf{RINN} demonstrated superior performance. In evaluations without phase data, the RRMSE metric still achieved results comparable to those obtained by other schemes when processing phase data. This makes the \textbf{RINN} a promising candidate for application on widely available IoT devices.}
    \label{tab:comparison}
    \begin{tabular}{ccccccccc}
    \toprule
    \multirow{2}{*}{Methods} & \multirow{2}{*}{\makecell{Applicable to\\ phaseless data}} & \multirow{2}{*}{\makecell{Applicable to one\\ sample data}} & \multicolumn{3}{c}{Results with phase data} & \multicolumn{3}{c}{Results with phaseless data} \\
    \cline{4-6} \cline{7-9}
    & & & SSIM & PSNR & RRMSE & SSIM & PSNR & RRMSE \\
\midrule
BP & \XSolidBrush & \CheckmarkBold & 0.78 & 21.11 & 0.16 & - & - & - \\
GS SoM & \XSolidBrush & \CheckmarkBold & 0.83 & 23.97 & 0.11 & - & - & - \\
CS-Net & \XSolidBrush & \XSolidBrush & 0.80 & 24.88 & 0.16 & - & - & - \\
DConvNet & \XSolidBrush & \XSolidBrush & 0.88 & 25.31 & 0.10 & - & - & - \\
Physics-Net & \XSolidBrush & \XSolidBrush & 0.92 & 26.05 & 0.08 & - & - & - \\ \hline
\textbf{RINN} & \CheckmarkBold & \CheckmarkBold & 0.91 & 28.34 & 0.08 & 0.68 & 13.4 & 0.11 \\
\bottomrule
\end{tabular}
\end{table}

\begin{figure}
    \centering
    \includegraphics[width=\linewidth]{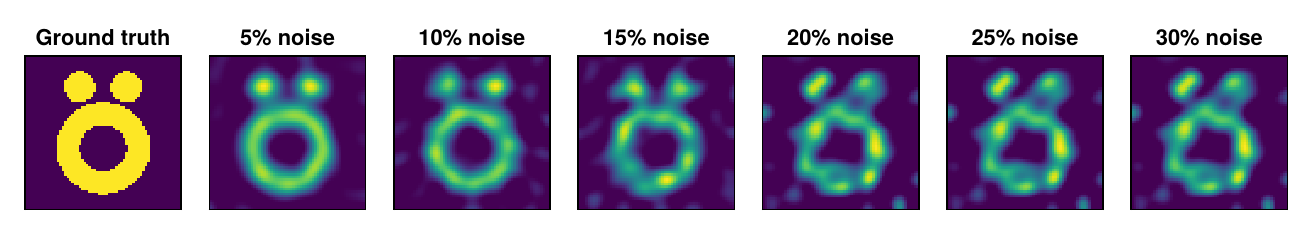}
    \caption{The processing results of the \textbf{RINN} network for noisy data. We use Austrian patterns as targets, then add different levels of noise to the amplitude of the received signals, and perform imaging using the \textbf{RINN} network. The results show that for lower levels of noise, the \textbf{RINN} imaging results are good; for higher levels of noise, although distorted, the basic pattern can still be discerned. This has potential to leverage the ubiquitous nature of RF signals in applications such as multimodal sensing.}
    \label{fig:noise}
\end{figure}

\textit{Baseline}. We maintain the same settings as in previous studies. We compare our method with two classical methods and three deep learning-based methods. The classical algorithms are the Backpropagation algorithm (BP)~\cite{belkebir2005superresolution} and the Subspace Optimization Method (GS SoM)~\cite{chen2009subspace}, while the three deep learning schemes are CS-Net~\cite{sanghvi2019embedding} (a scheme combining subspace optimization with CNN), DConvNet~\cite{yaoEnhancedDeepLearning2020} (a scheme based on the SegNet framework), and Physics-Net~\cite{liu2022physics} (a scheme embedding physical information into a CNN network). Since the aforementioned schemes are only applicable to data with phase, whereas our proposed \textbf{RINN} is compatible with phaseless data, we conducted a comparison of basic structures using data with phase~\footnote{When testing with phase data, we set the scattering loss $\mathcal{L}_{sa}$ of \textbf{RINN} to the mean square error of the scattered wave.} and also report the results of the \textbf{RINN} model on phaseless data.

\textit{Evaluation Metrics}. We used three evaluation metrics to evaluate the performance of t
\textbf{RINN}: structural similarity index measure (SSIM), peak signal-to-noise ratio (PSNR), and relative root mean square error (RRMSE). The SSIM is a measure of the similarity between two images, ranging from -1 to 1, with 1 indicating that the two images are identical. The PSNR is a measure of the quality of the image, with higher values indicating better quality. The RRMSE is a measure of the error between the predicted image and the ground truth image, with lower values indicating better results.

We initially conducted a qualitative assessment of RINN's imaging capabilities for targets of various shapes, followed by a comprehensive quantitative comparison of its performance with classical methods. Finally, we demonstrated its robustness to noisy data.

\subsection{Results}
\textit{Imaging results of phaseless data}. We first verified the ability of the \textbf{RINN} model to image the sensing region \( \mathcal{D} \) based on phaseless data. The results are shown in Fig.~\ref{fig:step}, where the top image is the ground truth and the bottom image is the output of \textbf{RINN}. Except for the ``Austria" pattern (the leftmost column), we randomly rotated four digit patterns from the MNIST dataset: 3, 8, 2, and 9. 
For the ``Austria" pattern, digit 3 pattern, and digit 8 pattern, the permittivity was set to 1.6; for digit 2 pattern, it was set to a smaller value of 1.1 (close to air's permittivity); and for digit 9 pattern, it was set to a larger value of 1.8. The results show that for different patterns and targets with varying permittivities, \textbf{RINN} produces clear imaging results and accurately captures target shapes.

\textit{Comparison results.} Table \ref{tab:comparison} shows the comparison results of the baseline. It includes six different algorithms, including two classical algorithms (BP and GS SoM), three deep learning schemes (CS-Net, DConvNet, and Physics-Net), and the proposed \textbf{RINN} in this paper. Note that the three classical learning schemes are designed based on phased data and require a dataset with multiple samples for training; while classical algorithms can obtain results through multiple iterations based on a one sample, they are still not suitable for phaseless data. The \textbf{RINN} model designed in this paper does not rely on large-scale datasets and is compatible with phaseless data. For imaging results with phased data, \textbf{RINN} demonstrates superior performance compared to other algorithms, such as having the best PSNR and RRMSE. For phaseless data, the lack of phase information leads to some degradation in imaging quality, but the RRMSE metric (0.11) remains close to the results obtained by other algorithms using phased data, indicating that \textbf{RINN} is particularly promising for deployment in radio frequency sensing applications.

\textit{The results of noisy data}. Fig.~\ref{fig:noise} shows the imaging results of \textbf{RINN} based on noisy phaseless data. The scatterer uses an Austrian pattern. We added varying levels of Gaussian noise to the amplitude data of the received signal and then used the \textbf{RINN} network for imaging. The results indicate that for lower levels of noise (less than 15\%), the imaging results are good. For higher levels of noise, the imaging patterns are somewhat distorted, but the basic pattern can still be vaguely discerned. We believe this still has potential to provide advantages in non-line-of-sight sensing using RF signals in multi-modal sensing systems compared to traditional modalities such as vision.

\section{Conclusion}
\label{sec:conclusion}
This article introduces \textbf{RINN}, a network based on one sample and phaseless data for RF imaging. Leveraging the idea of PINN, we embed physical information into neural networks, where the presence of physical constraints eliminates the need for large datasets during network training. We have modified the data processing and loss functions of the system to make it suitable for low-quality data (containing noise and no-phase) provided by a wide range of RF devices. Our numerical evaluation results show that \textbf{RINN} outperforms 5 classical algorithms in imaging with phase data (RRMSE is 0.08), moreover, \textbf{RINN} is also compatible with phaseless data (RRMSE is 0.11). These characteristics of \textbf{RINN} provide potential deployment possibilities for EISP technology on widely used RF devices.

\bibliographystyle{unsrtnat}
\bibliography{zotero,ref}  






\end{document}